\documentclass{article}
\usepackage{epsfig}
\usepackage{amssymb}
\usepackage{amsmath}
\usepackage{amsfonts}
\newcommand \be{\begin{eqnarray}}
\newcommand \ee{\end{eqnarray}}
\begin{document}
\begin{center}
{\bf Two Particles in a Trap }\\
\bigskip
\bigskip
H. S. K\"ohler \footnote{e-mail: kohlers@u.arizona.edu} \\
{\em Physics Department, University of Arizona, Tucson, Arizona
85721,USA}\\
\end{center}
\date{\today}

\begin{abstract}
The Busch-formula relates the energy-spectrum of two point-like particles 
interacting in a 3-D isotropic Harmonic Oscillator trap to the free 
scattering phase-shifts of the particles. 
This formula is used to find an expression for the \it shift \rm in the spectrum
from the unperturbed (non-interacting) spectrum rather than the spectrum
itself. 
This shift is shown to be approximately $\Delta=-\delta(k)/\pi\times dE$, 
where $dE$ is the spacing between unperturbed energy levels. 
The  resulting difference from the Busch-formula is typically $<\frac{1}{2}\%$
except for the lowest energy-state and small scattering length when it is
$3\%$. It goes to zero when the scattering length $\rightarrow \pm \infty$.

The   energy shift $\Delta$  is familiar from a related 
problem, that of two particles in a spherical infinite square-well trap  
of radius $R$ in the limit $R\rightarrow \infty$. The approximation is
however as large as $30 \%$ for finite values of $R$, a situation 
quite different from the Harmonic Oscillator case.

The square-well  results for $R\rightarrow \infty$ led to the use of
in-medium (effective) interactions in nuclear matter calculations that were
$\propto \Delta$  and known as the \it phase shift approximation \rm.
Our results indicate that the validity of this approximation depends on the trap
itself, a problem already  discussed by DeWitt more than 50 years ago for a 
cubical vs spherical trap. 
\end{abstract}

\section{Introduction}
In some of his early attempts of developping a many-body theory of nuclei,
Brueckner assumed the in-medium (effective) two-body interaction to be
$\propto \tan\delta(k)$, $\delta(k)$ being the scattering phase-shift as a
function of relative momentum $k$.
The initial problem studied was that of an 'infinite' system of nuclear
matter for the purpose of calculating binding energy and saturation
properties. The $\tan\delta(k)$ approximation came from the assumption that
the Reactance matrix, a part of scattering theory, could be used as an
in-medium interaction in the many-body problem. (The diagonal part of this
matrix is $\propto \tan\delta(k)$). This implies that a 
principal value Green's function 
with an integration over a semi-continuous spectrum  would be justified in
this case with the box of nuclear matter assumed to  be very and even
'infinitely' large. 
This assumption was substantiated in a paper by  Reifman and DeWitt\cite{rei56}.
It was however soon realised to be incorrect. Several authors (one of them
DeWitt) showed that the correct limit for two particles in a big box would 
yield an in-medium interaction $\propto \delta(k)$ rather than
$\tan\delta(k)$. 
\cite{fuk56,dew56,rie56}
\footnote{For small $\delta$ e.g. at low density this might not make any difference
but for problems of interest to-day with $\delta\approx \pi/2$ it obviously
would.}
The Reactance matrix is part of scattering theory. The nuclear matter
problem assumes a box, although large but still finite  with boundary 
conditions different from that of scattering theory. A particle in this box 
has a discrete rather than a continuous spectrum and that makes a difference..

The proofs in the referred papers differ but the essential point is that
the integration over the continuous spectrum vs the summation over the discrete
spectrum differs even though the level-density in a big box goes to zero as
the box-size increases. The principal value integration relates to the
scattering  problem  with
boundary conditions different from that for a box where the wave-functions
are zero at the edge of the box even in the limit of 'infinitely' large. 

The exact statement of the results in the referenced papers is that the
energy-shift $\Delta$ due to the interaction of two-particles confined 
in a spherical box in the limit when the size of the box approaches infinity is
given by $\Delta=S\times dE$
where $dE$ is the spacing between the levels of the unperturbed spectrum
and
\footnote{It was pointed out by DeWitt\cite{dew56}, after a comment by 
Brueckner that the result may be different for a box other than spherical. }
\begin{equation}
S=-\frac{\delta(k)}{\pi}.
\label{d}
\end{equation}
In nuclear matter studies this has been  referred to as the 'phase-shift
approximation' a term adopted here. 
It was for example used in early calculations on the
neutron-gas\cite{bru60,soo60} and used as a first approximation for nuclear
matter studies.\cite{hsk83}

The question of the author arose whether the result above might be more
general.
The problem of two particles trapped in a potential well is of interest for
atomic as well as nuclear physics  studies. The Busch-formula\cite{bus98}
relates the \it energy-spectra \rm of the two particles in a Harmonic Oscillator
well to the scattering phase-shifts for point-like potentials. 
This formula is here
rewritten in terms of the \it energy-shifts \rm rather than energy.  
Results below show that the phase-shift approximation for these shifts 
is (surprisingly) in practically exact agreement with the Busch-formula. 
The largest difference is found for the lowest energy-state ($n=0$) and small
scattering length but it decreases rapidly  for $n>0$
and larger scattering lengths.
For comparison are also shown results for the spherical infinite square well
as a function of the radius of the sphere. Although the approximation
becomes exact in the limit $R\rightarrow \infty$, it is in general
much worse in this case than for the H.O.

\section{3-D Harmonic Oscillator Well}
With the energy in units of $\hbar\omega$ 
 the total energy for two non-interacting particles with zero angular
 momentum in a 3-D Harmonic Oscillator well is $$E_{tot}=E+E_{cm}=2n+3$$
where $E_{cm}=\frac{3}{2}$ is the center-of mass energy .
It is convenient to choose $$a_{osc}=\sqrt{\hbar/m\omega}$$ as the unit of
length. With $\eta=2E$ the Busch-formula reads \cite{jon02}
\begin{equation}
\tan\delta(k)=-\frac{\sqrt{\eta}\Gamma[(1-\eta)/4]}{2\Gamma[(3-\eta)/4]}
\label{busch}
\end{equation}
where $k=\sqrt{\frac{\eta}{2}}/a_{osc}$.

The level-spacing in the uncorrelated system is $dE=2$ and following the
notation above  one finds the level-shift to be $\Delta=2S$. 
In the case of the spherical box
above $S$ is given by eq. (\ref{d}). It is the purpose of the present work to
find $S$ for the 3-D H.O. with the Busch spectrum.
One finds $$\eta=4n+3+4S$$.
Using the reflection formula for $\Gamma$-functions 
\begin{equation}
\Gamma(1-x)=\frac{\pi}{\Gamma(x)\sin(\pi x)}
\label{refl}
\end{equation}
one finds

\begin{equation}
\tan \delta(k)=-A(z)\tan(S\pi)
\label{busch2}
\end{equation}
where

\begin{equation}
A(z)=\frac{\sqrt{z-\frac{1}{4}} \Gamma(z)}{\Gamma(z+\frac{1}{2})}
\label{busch3}
\end{equation}
 
with z=n+S+1. 

The beta-function $B(x,y)$ is defined by
\begin{equation}
B(x,y)=\frac{\Gamma(x)\Gamma(y)}{\Gamma(x+y)}
\label{busch4}
\end{equation}

and for $y$ fixed and $x$ large one has \footnote{I owe my thanks to Dr Jerry
Yang for showing me this relation on Wikipedia}.
\begin{equation}
B(x,y)\approx \Gamma(y)x^{-y}
\label{busch5}
\end{equation}

so that  for $z$ large one would have
$$\frac{\Gamma(z+\frac{1}{2})}{\Gamma(z)} \approx \sqrt{z}$$
but in eq. (\ref{busch3}) one has the factor $$\sqrt{z-\frac{1}{4}}$$.

Numerical tests show that a better approximation  indeed is
$$\frac{\Gamma(z+\frac{1}{2})}{\Gamma(z)} \approx \sqrt{z-\frac{1}{4}}$$

\begin{figure}
\centerline{
\psfig{figure=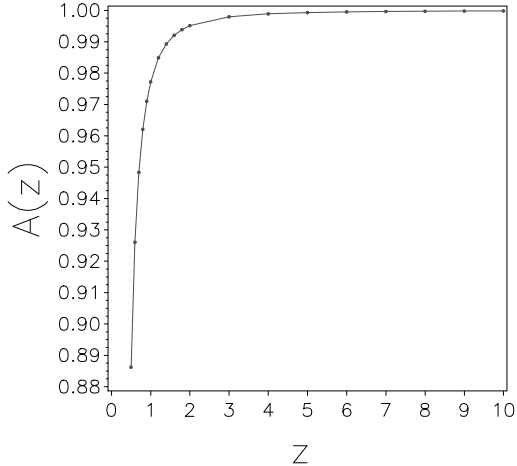,width=7cm,angle=0}
}
\vspace{.0in}
\caption{
The function $A(z)$ defined in the text.
}
\label{bush1}
\end{figure}

One finds that the  values of $z$ that is needed in the Busch-formula is
$z\geq \frac{1}{2}$ and
Fig. \ref{bush1} shows the function $A(z)$ for a useful range of $z$.

One finds $A \rightarrow 1$ quite rapidly with $z$ (and consequently with $n$) 
so that in this limit eq. (\ref{busch2}) yields  
\begin{equation}
S \rightarrow-\frac{\delta(k)}{\pi}
\label{ho}
\end{equation}
which is the phase-shift approximation given in eq. (\ref{d}).

\begin{figure}
\centerline{
\psfig{figure=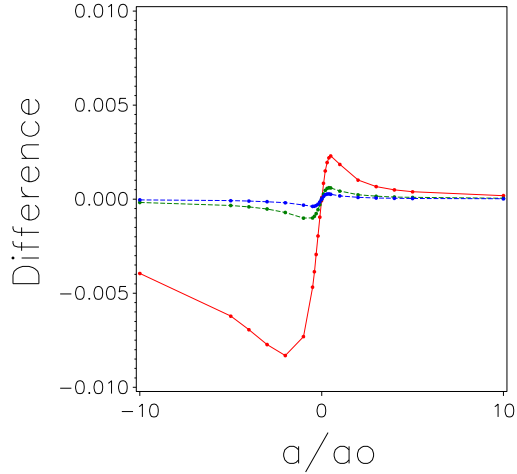,width=7cm,angle=0}
}
\vspace{.0in}
\caption{
The difference between the energy-shifts calculated with the Busch-formula
and that obtained in the phase-shift approximation for $n=0,1,2$. The largest
difference is as expected for $n=0$ (red on-line), while it is appreciably
smaller for $n=1$ (green on-line) and $n=2$ (blue on-line). The smallest value of
$z$ in the function $A(z)$ is $\frac{1}{2}$ (see Fig. \ref{bush1}), which
occurs for $n=0$ and $\delta=-\frac{\pi}{2}$. The largest difference shown
is consequently for this value of $n$ and for $a<0$ where the shift is
negative. 
}
\label{bush2}
\end{figure}

\begin{figure}
\centerline{
\psfig{figure=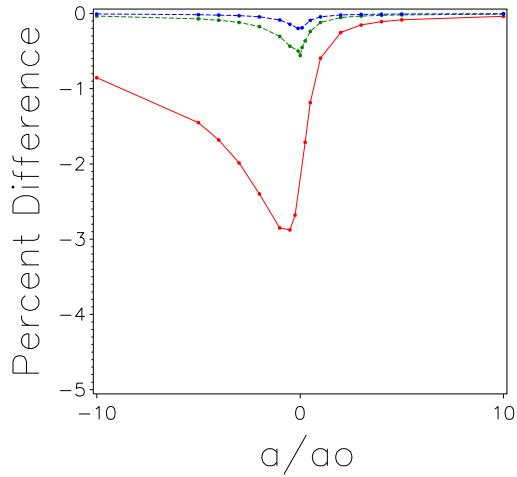,width=7cm,angle=0}
}
\vspace{.0in}
\caption{
Similar to Fig. \ref{bush2} but the difference is here shown as the percentage of
the energy-shift.  The maximum difference is $\approx -3\%$
(for $n=0$, red on-line) but a rapid decrease with increasing 
$n$ is seen and is less 
than $\frac{1}{2} \%$ for $n=1$ (green on-line) and even smaller for
$n=2$ (blue on-line).
}
\label{bush3}
\end{figure}

A correction $dS$ to $S$ due to $A(z)\neq 1$ is obtained from
\begin{equation}
dS\approx
-\frac{1}{\pi}(1-A^{-1}(z))\frac{\tan{\delta(k)}}{1+\tan^2\delta(k)}
\label{correct}
\end{equation}
This shows a dampening of the correction with increasing value of
$\tan^2{\delta}$. In the limit when $\delta=\frac{\pi}{2}$ the correction goes
to zero. The energy-shift is then exactly given by
$\frac{1}{\pi}\frac{\pi}{2}=\frac{1}{2}$, a well-known
result.\cite{bus98,jon02}
Fig. \ref{bush2} shows the difference between the energy shifts calculated
by the Busch-formula and the phase-shift-approximation for quantum levels
$n=0,1$ and $2$ as a function of scattering length. It is seen to
decrease rapidly with $n$ and also with $\pm a$. The peaks are a result 
of the competition
between the $A(z)$ and $\tan{\delta}$ corrections respectively shown in eq.
(\ref{correct}). 
Fig. \ref{bush3} shows the same differences but in $\%$ of the Busch result
for $n=1,2,3$ again showing the rapid decrease with increasing values of
$n$ and with $\pm a$. 

\section{Spherical Square Well}
The problem of two particles in a spherical box was treated by several
authors referred to in the Introduction. 
In these several works the emphasis was on the energy-shift in the 
limit of the radius of the sphere going to 'infinity' with the result
given by eq.  (\ref{d}), the phase-shift approximation.
For comparison with the 3-D H.O. it is of interest to also display the
shift as a function of the ratio of scattering length to radius and
as a function of  quantum-numbers $n$ of the spherical box. 

The simplest solution \cite{fuk56,got} of the problem at hand is to 
explicitly consider the wave-functions of the two particles 
in this box.
Considering only $s$-states the radial wavefunctions of free non-interacting
particles are
\begin{equation}
\Psi(r) \propto \frac{1}{r}\sin(k^{(0)}r)
\label{ucorr}
\end{equation}
With  two-body interaction 
the wave-function outside the range of the two-body interaction is
\begin{equation}
\Psi(r)\propto \frac{1}{r}\sin(kr+\delta(k))
\label{corr}
\end{equation}

The boundary condition implies that the wave-functions vanish at the
boundary of the sphere assumed to have a radius $R$. (One immediately sees
the simplicity of the formalism relative to that of the 3-D H.O. or a
cubical box.)
Thus, for the non-interacting case 
\begin{equation}
k^{(0)}_nR=n\pi
\label{ucorr1}
\end{equation}
and for the interacting case
\begin{equation}
k_nR+\delta(k_n)=n\pi
\label{corr1}
\end{equation}

The spacing between unperturbed levels is
$$dE={{k^{(0)}_{n+1}}^2-{k^{(0)}_n}^2}=
(k^{(0)}_{n+1}+k^{(0)}_n)\left(\frac{\pi}{R}\right)$$

The energy shift
$$\Delta=k_n^2-{k^{(0)}_n}^2=-(k_n+k_n^{(0)})\left( \frac{\delta(k_n)}{R}\right)$$ 
divided by the unperturbed energyspacing, i.e. $S$, is then

\begin{equation}
S=-\frac{k_n+k^{(0)}_n}{k^{(0)}_{n+1}+k^{(0)}_n}
\left(\frac{\delta(k_n)}{\pi}\right )
\label{S}
\end{equation}
 
The energy-shifts and -spacings decrease with $R$ and $n$ $\rightarrow
\infty$ and then
$$S\rightarrow -\frac{\delta(k_n)}{\pi}$$ as before.

The phase-shift approximation was tested for convergence as a function of
box radius $R$ and quantum number $n$. For this purpose the 
expression for $S$ was rewritten as follows
\begin{equation}
S=\frac{1}{\pi}\left(\frac{\delta(k_n)-2n\pi)}{2n}\right)\frac{\delta(k_n)}{\pi}
\label{sphe1}
\end{equation}

The radius $R$ only appears implicitly through the relation
$$k_n=(n\pi-\delta(k_n))/R$$
to be solved selfconsistently.

In the (unphysical) case that $\delta$ would be independent of $k$ one
would have the situation where $S$ is independent of $R$, but only of $n$
and of course of $\delta$.

\begin{figure}
\centerline{
\psfig{figure=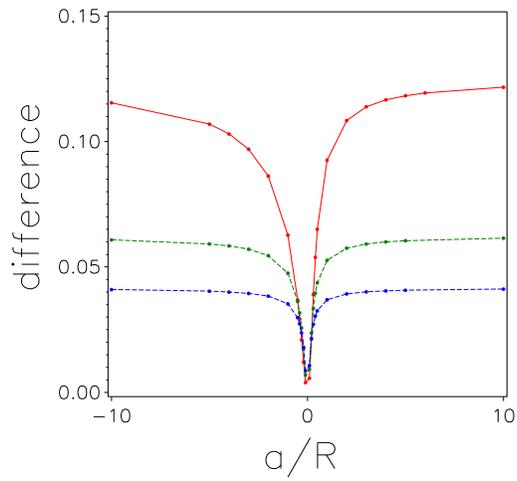,width=7cm,angle=0}
}
\vspace{.0in}
\caption{
The difference between the exact and phase-shift approximation for 
two-particles in the spherical trap for three values of the quantum-number
$n=1$,  (red on-line), $n=2$ (green on-line) and $n=3$ (blue on-line).
Notice the difference with the similar plot in Fig. \ref{bush2} for the
H.O. trap.
}
\label{sphere1}
\end{figure}

\begin{figure}
\centerline{
\psfig{figure=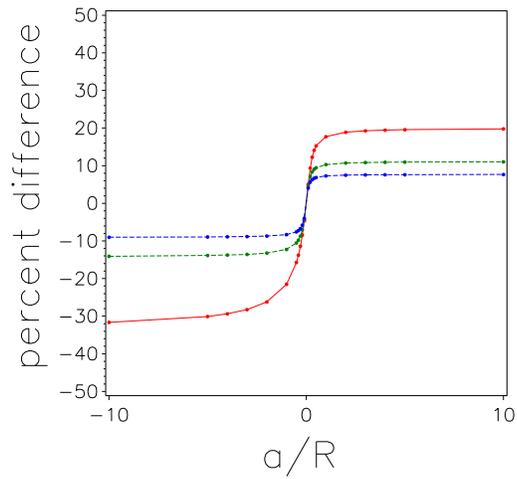,width=7cm,angle=0}
}
\vspace{.0in}
\caption{
Similar to Fig. \ref{sphere1} but the difference is shown as percentage of
the energy-shift.
}
\label{sphere2}
\end{figure}

Fig. \ref{sphere1} shows the difference between the 'exact' shift and the
phase-shift approximation $-\frac{\delta}{\pi}$ of eq. (\ref{d}), while
Fig. \ref{sphere2} shows the difference in $\%$ of the exact shift. 

Compared with the analogous result for the H.O. there are notable
differences. While the  largest deviation from the phase-shift approximation
is $3\%$ (and typically much smaller) in the H.O. case, it can for the
sphere be even as large as $30\%$. But even though the devation from the
phase-shift approximation in the H.O. is small it has its maximum for large
values of $a_{osc}$ (or small scattering lengths)
while going to zero for small values (or large
scattering lengths), which is opposite to the situation encoutered 
for the spherical well. 
Both Figs. \ref{sphere1} and \ref{sphere2} illustrate however the known
result that $S\rightarrow -\frac{\delta(k)}{pi}$ as $R \rightarrow \infty$
as shown in previous works already referred to above.\cite{fuk56,dew56,rie56}.

\section{Summary}
The phase-shift approximation $-\frac{1}{\pi}\delta(k)dE$ for the energy-shift
due to the interaction of two partcles in a trap, where $\delta(k)$ is the
free particle scattering phase-shift and  dE  the energy-level spacing 
of the unperturbed spectrum,  has been  investigated. 

The traps considered were a 3-D Harmonic Oscillator and a spherical 
infinite square well.  The square-well  case was treated a long time ago
as referenced above, with the main emphasis on the large size limit of the 
trap with the result expressed by eq. (\ref{d}). 
Somewhat of a surprise was the finding of the present investigation 
that a not only  somewhat similar situation exists for the H.O. trap but that 
the approximation is so much more accurate for this trap.  

The differences between the results
for the two respective traps are seen by comparing Figs \ref{bush2}  
and \ref{bush3} for the H.O.trap with the Figs. \ref{sphere1} and 
\ref{sphere2} for the square-well trap.  While in the  square well
case, the differences (errors) by using the phase-shift approximation are as
large as $30\%$, the differences are much smaller and even practically zero
in the H.O. case.  With $a\rightarrow \pm \infty$ the
phase-shift approximation becomes exact in the H.O. case while in the 
spherical case the error increases. While Figs \ref{sphere1} and
\ref{sphere2} show that the approximation becomes perfect
when $a/R$ goes to zero (or the radius of the square well goes to
$\infty$) that is when (unexpectedly) the largest correction,
although still small, is found in the H.O.  case.

In all the work above like in the Busch-formula,  it is assumed that 
the range of the potential is
small compared to the size of the trap. Considerations of range corrections
are beyond the scope of this investigation. It has been the subject of a
recent publication.\cite{luu10}

The question arises of course whether the phase-shift approximation has a more
general validity e.g. for the shifts in a cubical box, as derived  by
L\"uscher.\cite{lu91} It is however already seen that the situation is
rather different in the 3-D H.O. case compared to the case of a spherical
well. The conceptual differences  between scatterings and interactions in
the spherical vs cubical box was already pointed out by DeWitt in a 
detailed discussion of this matter.\cite{dew56}. 
\footnote{Quote: Only the spherical box (with spherical waves) is suitable
for establishing a connection between single scattering processes and
discrete spectrum theory....}

\end{document}